\RequirePackage{lineno}
\documentclass[onecolumn,aps,prc,superscriptaddress]{revtex4}
\usepackage{epsfig,graphics}
\usepackage{graphicx}
\usepackage{subfigure}
\usepackage{dcolumn}
\usepackage{bm}
\usepackage[usenames]{color}
\usepackage{ulem} 
\usepackage{url}
\usepackage{amsmath}
\usepackage{booktabs}
\usepackage{multirow}
\begin{document}
\title{Multiparticle azimuthal cumulants from transverse momentum conservation and collective flow}

\author{Mu-Ting Xie}
\author{Guo-Liang Ma}
\email{glma@fudan.edu.cn}
\affiliation{Key Laboratory of Nuclear Physics and Ion-beam Application (MOE), Institute of Modern Physics, Fudan University, Shanghai 200433, China}
\affiliation{Shanghai Research Center for Theoretical Nuclear Physics, NSFC and Fudan University, Shanghai $200438$, China}
\author{Adam Bzdak}
\email{bzdak@fis.agh.edu.pl}
\affiliation{AGH University of Science and Technology,\\
Faculty of Physics and Applied Computer Science,
30-059 Krak\'ow, Poland}


\begin{abstract}
We calculate the $n$th order of $2k$-particle azimuthal cumulants $c_n\{2k\}$ based on transverse momentum conservation (TMC) and collective flow $v_n$(n=2,3). We demonstrate that the TMC effect only leads to a nonzero $c_n\{2k\}$ with the sign of $(-1)^{nk}$ and the magnitude inversely proportional to $(N-2k)^{nk}$. The interplay between TMC and collective flow can change the signs of $c_2\{4\}$, $c_3\{2\}$ and $c_3\{4\}$ at some values of multiplicity $N$, which could provide a good probe to study the onset of collectivity and search for the substructure of proton in small colliding systems.

\end{abstract}

\maketitle

\section{Introduction}

Quarks and gluons which are confined inside nucleons can be released under an environment with high temperature and (or) large baryon number chemical potential~\cite{Bzdak:2019pkr, deForcrand:2002hgr,Ding:2015ona,Aoki:2006we,Ejiri:2008xt}. A large amount of experimental results from both the Relativistic Heavy Ion Collider (RHIC) and the Large Hadron Collider (LHC) have shown that a new deconfined QCD matter, so-called strongly coupled quark gluon plasma (sQGP), has been produced in the early stage of high energy nucleus-nucleus (A+A) collisions~\cite{Adams:2005dq, Adcox:2004mh,Aamodt:2008zz}. Among the most important experimental evidences is the observation of strong collective flow of the produced particles in A+A collisions. This is because the collective flow is considered to result from the collective expansion of sQGP, which can transfer the asymmetry of the initial space geometry into the anisotropy of the final particles' momenta~\cite{Ollitrault:1992bk,Poskanzer:1998yz,Adams:2004bi,ALICE:2011ab,ATLAS:2012at}.

Compared with the large A + A colliding systems, proton-proton or proton-nucleus collisions are called small colliding systems. Recent experimental results surprisingly show that the small colliding systems carry ``collective flow'' as strong as that in large colliding systems, which poses a challenge to our current understanding of the strong collective flow due to the expansion of sQGP~\cite{Dusling:2015gta,Loizides:2016tew,Nagle:2018nvi}. In order to understand the origin(s) of collective flow in small colliding systems, lots of theoretical efforts have been made, which basically can be divided into two categories according to whether the origin comes from the final or initial state. For example, hydrodynamics can transform the initial geometric asymmetry into the final momentum anisotropic flow through pressure gradient of the QGP, which can well describe the experimental results~\cite{ Bozek:2011if,Bzdak:2013zma,Shuryak:2013ke,Qin:2013bha,Bozek:2013uha,Bozek:2015swa,Song:2017wtw}. Parton cascade could  achieve a similar conversion through an escape mechanism~\cite{ Ma:2014pva,Bzdak:2014dia,He:2015hfa,Lin:2015ucn,Ma:2021ror}. On the other hand, the initial state of color glass condensate (CGC) has also been proposed as a possible mechanism, contributing to the experimentally measured correlations in small colliding systems~\cite{Dumitru:2010iy,Dusling:2013oia,Skokov:2014tka,Schenke:2015aqa,Schlichting:2016sqo,Kovner:2016jfp,Iancu:2017fzn,Mace:2018vwq,Nie:2019swk}. Interestingly, using the CGC effective field theory coupled to hydrodynamical simulations, it has been found that fluctuating substructure of proton must be included in order to better match the experimentally observed anisotropic flow in small systems~\cite{Mantysaari:2016ykx,Mantysaari:2020axf,Schenke:2021mxx}. A multiphase transport model with sub-nucleon geometry can better describe the multiplicity dependence of $c_n\{4\}$, which demonstrates the importance of incorporating the substructure of proton in studies of small colliding systems~\cite{Zhao:2021bef}.

On the other hand, experimentalists have made lots of efforts to measure collective flow in small colliding systems using different observables. The multi-particle azimuthal cumulant has been proposed as an advanced and powerful tool to explore the collectivity of many-body systems, because it can effectively reduce few-body non-flow contribution~\cite{Borghini:2000sa}. The $n$th order of 2$k$-particle azimuthal cumulant $c_n\{2k\}$ is defined as follows,
\begin{eqnarray} \begin{aligned}
\label{eq:cn2468}
&c_n\{2\} = \langle e^{in(\phi_1-\phi_{2})}\rangle \\
&c_n\{4\} = \langle e^{in(\phi_1+\phi_2-\phi_3-\phi_{4})}\rangle - 2\langle e^{in(\phi_1-\phi_{2})}\rangle^2 \\
&c_n\{6\} = \langle e^{in(\phi_1+\phi_2+\phi_3-\phi_{4}-\phi_{5}-\phi_{6})}\rangle - 9\langle e^{in(\phi_1+\phi_2-\phi_3-\phi_{4})}\rangle\langle e^{in(\phi_1-\phi_{2})}\rangle \\
&+12\langle e^{in(\phi_1-\phi_{2})}\rangle^3 \\
&c_n\{8\} = \langle e^{in(\phi_1+\phi_2+\phi_3+\phi_{4}-\phi_{5}-\phi_{6}-\phi_{7}-\phi_{8})}\rangle - 16\langle e^{in(\phi_1+\phi_2+\phi_3-\phi_{4}-\phi_{5}-\phi_{6})}\rangle\langle e^{in(\phi_1-\phi_{2})}\rangle\\
&-18\langle e^{in(\phi_1+\phi_2-\phi_3-\phi_{4})}\rangle^2+144\langle e^{in(\phi_1+\phi_2-\phi_3-\phi_{4})}\rangle\langle e^{in(\phi_1-\phi_{2})}\rangle^2-144\langle e^{in(\phi_1-\phi_{2})}\rangle^4 \\
\end{aligned} \end{eqnarray}
where $\phi_i$ is the azimuthal angle of $i$th particle, n=2 and 3 correspond to the elliptic and triangular flow, respectively. The experimental measurement has shown that  4,6,8-particle elliptic flow cumulants are almost same in p+p and p+Pb collisions, which indicates the existence of multi-particle correlations in small colliding systems~\cite{Khachatryan:2015waa}. The recent experimental results show four-particle azimuthal elliptic flow cumulant $c_2\{4\}$ changes its sign, from positive to negative, as the multiplicity increases~\cite{Khachatryan:2016txc,Aaboud:2017blb}, which could be related to the onset of collectivity in small colliding systems~\cite{Zhao:2020pty}.

The conservation laws are also an obvious source of the azimuthal correlation between particles,
see, e.g., \cite{Borghini:2000cm,Borghini:2002mv,Chajecki:2008vg,Chajecki:2008yi,Pratt:2010zn,Bzdak:2010fd,Borghini:2003ur}. For instance, transverse momentum conservation (TMC) is an important background for the experimental measurement of directed flow ($v_1$), especially in peripheral A+A collisions~\cite{ATLAS:2012at,Alt:2003ab}, mainly because the $1/N$ correction caused by TMC is very large. Our recent studies have found that TMC is also important for understanding of the behavior of elliptic flow in small colliding systems. We found that with the increasing value of the total multiplicity $N$, the TMC effect leads to a positive $2k$-particle elliptic flow cumulants $c_2\{2k\}$ which satisfies the dependence of $1/(N-2k)^{2k}$~\cite{Bzdak:2017zok}. The four-particle elliptic flow coefficient $c_2\{4\}$ will change its sign if TMC interplays with hydro-like elliptic flow, which can naturally explain the observed behavior of the multiplicity dependence of $c_2\{4\}$ in the small colliding systems at the LHC~\cite{Bzdak:2018web}.

In this work, we will generalize the $2k$-particle flow cumulant $c_n\{2k\}$ induced by TMC to higher orders in Section~\ref{tmconly}. By including the interplay between TMC with hydro-like elliptic and triangular flow, we will focus on the third order of multi-particle azimuthal cumulants $c_3\{2\}$ and $c_3\{4\}$ in Section~\ref{tmcflow} and compare them with the recent experimental data in small colliding systems from the LHC in Section~\ref{compare}. Finally, conclusions are given in Sec.~\ref{conclusions}.

\section{$c_n\{2k\}$ from transverse momentum conservation}
\label{tmconly}
The previous studies have shown that the cumulant flow coefficients can arise from transverse momentum conservation~\cite{Bzdak:2017zok}. Let us first briefly review our calculation method. To calculate the TMC contribution to $2k$-particle azimuthal cumulant $c_n\{2k\}$, the following term has to be first calculated:
\begin{equation}
\langle e^{in(\phi_1+...+\phi_k-\phi_{k+1}-...-\phi_{2k})}\rangle =\frac{\int_{\Omega}e^{in(\phi_1+...+\phi_k-\phi_{k+1}-...-\phi_{2k})}f(\vec{p}_1,\vec{p}_2,...,\vec{p}_{2k})p_1p_2...p_{2k}d\phi_1d\phi_2...d\phi_{2k}dp_1dp_2...dp_{2k}}
{\int_{\Omega}f(\vec{p}_1,\vec{p}_2,...,\vec{p}_{2k})p_1p_2...p_{2k}d\phi_1d\phi_2...d\phi_{2k}dp_1dp_2...dp_{2k}} \,,
\label{eq:mean1}
\end{equation}
where $p$ stands for a particle transverse momentum, and the integration is over a given acceptance phase-space region $\Omega$ of $2k$ particles. Note that we denote $e^{in(\phi_1+...+\phi_k-\phi_{k+1}-...-\phi_{2k})}$ as $e^{in(\phi_1+...-\phi_{2k})}$ in the following for simplicity. The $2k$-particle probability distribution $f(\vec{p}_{1},\vec{p}_{2}...\vec{p}_{2k})$ for the $N$-particle system under TMC can be approximately given by the central limit theorem, see, e.g., \cite{Borghini:2000cm,Chajecki:2008vg,Bzdak:2010fd}
\begin{equation}
f(\vec{p}_{1},...,\vec{p}_{2k})=f(\vec{p}_{1})\cdots f(\vec{p}_{2k})\frac{N%
}{N-2k}\exp \left( -\frac{(\vec{p}_{1}+...+\vec{p}_{2k})^{2}}{%
(N-2k)\left\langle p^{2}\right\rangle _{F}}\right) ,
\label{eq:2kp}
\end{equation}
where $N$ is the total number of particles in the system ($2k<N$) and $\langle p^{2}\rangle _{F}$ stands for mean value of squared $p$ in the full phase-space $F$,
\begin{equation}
\langle p^{2}\rangle _{F}=\frac{\int_{F}p^{2}f(\vec{p})d^{2}\vec{p}}{%
\int_{F}f(\vec{p})d^{2}\vec{p}}.
\end{equation}
The equation (\ref{eq:mean1}) contains an integration over transverse momenta. In our work, we calculate the cumulants at a given $p$, which substantially simplifies the problem. In this case Eq. (\ref{eq:mean1}) can be rewritten as
\begin{equation}
\langle e^{in(\phi_1+...-\phi_{2k})}\rangle|_{p_1,p_2...p_{2k}}=\frac{\int_{\Omega}e^{in(\phi_1+...-\phi_{2k})}\exp(X)d\phi_1d\phi_2...d\phi_{2k}}{\int_{\Omega}\exp(X)d\phi_1d\phi_2...d\phi_{2k}} \,,
\label{eq:mean2}
\end{equation}
where
\begin{equation}
X = -\frac{\sum_{0<l<j<2k+1}2p_lp_j\cos{(\phi_l-\phi_j)}}{(N-2k)\langle p^2\rangle_{F}} = -\frac{\sum_{l\neq j}p_lp_j e^{i(\phi_l-\phi_j)}}{(N-2k)\langle p^2\rangle_{F}} \,.
\label{eq:x}
\end{equation}
We can easily integrate the denominator of Eq. (\ref{eq:mean2}) which approximately equals to $(2\pi)^{2k}$ (here we keep the leading term, $e^{X} \approx 1$). However, the integration of the numerator cannot be achieved easily. But as $X$ scales inversely with $N-2k$, it can be presumed that $X$ is small enough so that an expansion in powers of $X$ can be employed if $N-2k$ is large enough. Then we can rewrite Eq. (\ref{eq:mean2}) as
\begin{equation}
\langle e^{in(\phi_1+...-\phi_{2k})}\rangle|_{p_1,p_2...p_{2k}}=\frac{\int_{\Omega}e^{in(\phi_1+...-\phi_{2k})}(1+X+\frac{X^2}{2}+\frac{X^3}{6}+...)d\phi_1d\phi_2...d\phi_{2k}}{(2\pi)^{2k}}.
\label{eq:exp1}
\end{equation}
With the knowledge of orthogonal functions, among the expansion terms $(1+X+\frac{X^2}{2}+\frac{X^3}{6}+...)$, only terms linearly correlated with $e^{-in(\phi_1+...-\phi_{2k})}$ can generate a nonzero outcome with a coefficient of $(2\pi)^{2k}$.
Here only $\frac{X^{m}}{m!}$ with $m\geq n k$ would contain terms linearly correlated with $e^{in(-\phi_1-...+\phi_{2k})}$. In our calculation we keep only the leading term coming from $\frac{X^{n k}}{(n k)!}$.

Observing Eqs. (\ref{eq:x}) and Eqs. (\ref{eq:exp1}), we can write the leading term as $C\frac{p_1^np_2^n...p_{2k}^n}{(N-2k)^{nk}\langle p^2\rangle^{nk}_F}$, where the coefficient $C$ can be calculated as follows. From Eqs. (\ref{eq:x}), we can see that among $X^{n k}$, each $X$ can provide a factor with a positive power $e^{i(\phi_l)}$ and a negative one $e^{-i(\phi_{j})},0<j\neq l<2k+1$. To reach $e^{in(-\phi_1-...-\phi_{k}+\phi_{k+1}+...+\phi_{2k})}$, the only possibility is that $n k$ positive power factors explicitly form the $e^{in(\phi_{k+1}+...+\phi_{2k})}$ and $n k$ negative power factors form the $e^{-in(\phi_{1}+...+\phi_{k})}$. For the positive power factors, there are $\frac{(n k)!}{(n!)^k}$ ways in total to form $e^{in(\phi_1+...+\phi_{k})}$. For the negative terms, $\frac{(n k)!}{(n!)^k}$ ways exist as well. Taking the coefficient coming from the exponential expansion $\frac{(-1)^{nk}}{(n k)!}$ into account, we obtain the 2$k$-particle azimuthal cumulant $c_n\{2k\}$ due to TMC with the leading term considered only, as shown below,
\begin{equation}
c_n\{2k\}=(-1)^{nk}\frac{(nk)!}{(n!)^{2k}}\frac{p_1^np_2^n...p_{2k}^n}{(N-2k)^{nk}\langle p^2\rangle^{nk}_{F}}.
\label{eq:cn2kTMC}
\end{equation}

Using the definitions of $c_n\{2k\}$ in Eqs. (\ref{eq:cn2468}), we summarize the TMC-induced
$c_n\{2k\}$ for $n=2,3,4$ and $2k=2,4,6,8$ in TABLE~\ref{table:TMC1}. From these results, we observe that except for the coefficients of $c_3\{2\}$ and $c_3\{6\}$ which are negative, the other coefficients are all positive. Moreover the magnitude of the TMC contribution to $c_n\{2k\}$ obviously decreases with the increasing value of $N$. Note that although our calculations concern only the leading term, these results can still be applied to describe the tendency and the magnitude of the TMC-induced $c_n\{2k\}$.

\begin{table}[htbp]
\caption{The $n$th order $2k$-particle cumulant $c_n\{2k\}$ from the global conservation of transverse momentum only.
}
\label{table:TMC1}
\centering
\begin{tabular}{p{30pt}p{80pt}p{80pt}p{80pt}p{80pt}}
\hline
\hline

$n$ &$c_n\{2\}$&$c_n\{4\}$&$c_n\{6\}$&$c_n\{8\}$\\
\hline
2&$\frac{1}{2}\frac{p_1^2p_2^2}{[(N-2)\langle p^2\rangle_{F}]^2}$&$\frac{p_1^2p_2^2p_3^2p_4^2}{[(N-4)\langle p^2\rangle_{F}]^4}$&$6\frac{p_1^2p_2^2...p_6^2}{[(N-6)\langle p^2\rangle_{F}]^6}$&$72\frac{p_1^2p_2^2...p_8^2}{[(N-8)\langle p^2\rangle_{F}]^8}$\\
\\
3&$-\frac{1}{6}\frac{p_1^3p_2^3}{[(N-2)\langle p^2\rangle_{F}]^3}$&$\frac{1}{2}\frac{p_1^3p_2^3p_3^3p_4^3}{[(N-4)\langle p^2\rangle_{F}]^6}$&$-7\frac{p_1^3p_2^3...p_6^3}{[(N-6)\langle p^2\rangle_{F}]^9}$&$261\frac{p_1^3p_2^3...p_8^3}{[(N-8)\langle p^2\rangle_{F}]^{12}}$\\
\\
4&$\frac{1}{24}\frac{p_1^4p_2^4}{[(N-2)\langle p^2\rangle_{F}]^4}$&$\frac{17}{144}\frac{p_1^4p_2^4p_3^4p_4^4}{[(N-4)\langle p^2\rangle_{F}]^8}$&$\frac{709}{288}\frac{p_1^4p_2^4...p_6^4}{[(N-6)\langle p^2\rangle_{F}]^{12}}$&$\frac{54193}{288}\frac{p_1^4p_2^4...p_8^4}{[(N-8)\langle p^2\rangle_{F}]^{16}}$\\
\hline
\hline
\end{tabular}
\end{table}

Our calculations are based on the assumption that X is small enough to make an expansion. As X scales inversely with $N-2k$, our calculations may tend to lose efficacy when $N-2k$ is very small, as higher orders terms need to be included. On the other hand, when $N-2k$ is very large, the TMC effects would be faint and the flow contribution would instead dominate the total outcome.

\section{$c_n\{2k\}$ from TRANSVERSE MOMENTUM CONSERVATION and flow}
\label{tmcflow}

In the previous section, the $2k$-particle cumulants $c_n\{2k\}$ from the transverse momentum conservation only have been calculated. In this section, the contribution coming from the collective flow will be included as well. The particle azimuthal distribution can be described as
\begin{equation}
\frac{dN}{d\phi}=\frac{g(p)}{2\pi}[1+\sum_{n} 2v_n(p)\cos(n(\phi-\Psi_n))],
\end{equation}
where $v_{n}$ is the $n$th order of the flow coefficient and $\Psi_n$ is the $n$th order event plane. Because the experimental results disclose that the directed flow $v_1$ and high orders of $v_n$ ($n>3$) are smaller than $v_2$, $v_3$, we only consider $v_2$ and $v_3$ in our calculations for simplicity.
After taking the collective flow into account, the $2k$-particle probability distribution of Eq.~(\ref{eq:2kp}) can be modified as (see, e.g., Refs. \cite{Bzdak:2010fd,Bzdak:2018web}),
\begin{equation}
f(\vec{p}_{1},...,\vec{p}_{2k})=f(\vec{p}_{1})\cdots f(\vec{p}_{2k})\frac{N%
}{N-2k}\exp\left(-\frac{({p}_{1x}+...+{p}_{2kx})^{2}}{%
2(N-2k)\langle p_x^{2}\rangle _{F}}-\frac{({p}_{1y}+...+{p}_{2ky})^{2}}{%
2(N-2k)\langle p_y^{2}\rangle _{F}} \right),
\end{equation}
 where
\begin{equation}
\langle p_x^{2}\rangle _{F}=\frac12\langle p^{2}\rangle _{F}(1+v_{2F})\\
\end{equation}
\begin{equation}
\langle p_y^{2}\rangle _{F}=\frac12\langle p^{2}\rangle _{F}(1-v_{2F})\\
\end{equation}
\begin{equation}
v_{2F}=\frac{\int_{F}v_2(p)g(p)p^2d^2p}{\int_{F}g(p)p^2d^2p}.
\end{equation}

Note that we set $\Psi_{2}=0$. With the help of Euler's formula, $\langle e^{in(\phi_1+...-\phi_{2k})}\rangle|_{p_1,...p_{2k}}$ is given by
\begin{equation}
\langle e^{in(\phi_1+...-\phi_{2k})}\rangle|_{p_1,...p_{2k}}=\frac{\int_{F}e^{in(\phi_1+...-\phi_{2k})}\exp(X)\displaystyle\prod_{j}[1+\sum_{m} v_m(p_{j})(e^{im(\phi_j+\Psi_m)}+e^{-im(\phi_j-\Psi_m)})]d\phi_1d\phi_2...d\phi_{2k}}{\int_{F}\exp(X)d\phi_1d\phi_2...d\phi_{2k}}
\label{integ}
\end{equation}
where
\begin{equation}
X=-\frac{({p}_{1x}+...+{p}_{2kx})^{2}}{%
2(N-2k)\langle p_x^{2}\rangle _{F}}-\frac{({p}_{1y}+...+{p}_{2ky})^{2}}{%
2(N-2k)\langle p_y^{2}\rangle _{F}}.
\end{equation}
This integral is not straightforward and generate a lot of terms. First, to simplify our consideration we assume that all the momenta are equal, i.e., $p_1=p_2=...=p_{2k}=p$. Next, as before, we expand $e^X$, i.e., $e^X=1+X+X^{2}/2!+...$, which allows to write the result as
\begin{equation}
\langle e^{in(\phi_1+...-\phi_{2k})}\rangle|_{p}=U_0+U_1Y+U_2\frac{Y^2}{2}+...+U_m\frac{Y^m}{m!}+...,
\end{equation}
where $Y = -\frac{p^2}{(N-2k)\langle p^2\rangle_{F}(1-v_{2F}^2)}$ and the coefficients $U_m$ depend on $v_2$, $v_3$, $p$, and $v_{2F}$. We include all the terms up to the one containing the pure TMC effect, i.e., $U_{nk}$ for $c_n\{2k\}$, and higher ones are neglected. The details of our calculation is shown in Appendix~\ref{appendix}.

We have obtained $c_2\{2\}|_p$ and $c_2\{4\}|_p$ in our previous work~\cite{Bzdak:2018web}. In this paper, we focus on $c_3\{2\}|_p$ and $c_3\{4\}|_p$. Given that $v_2=0.05,v_3=0.0175,v_{2F}=0.025$, the terms that are about 100 times (or more) smaller than the largest term in a given $U_n$ are omitted. The full results can be found in Appendix~\ref{appendix}. We give their approximate expressions below. The two-particle triangular cumulant coefficient with a momentum $p$, $c_3\{2\}|_p$, is given by
\begin{equation}
\begin{aligned}
&c_3\{2\}|_p = U_0+U_1Y+U_2\frac{Y^2}{2}+U_3\frac{Y^3}{6}, \quad ~Y=-\frac{p^2}{(N-2)\langle p^2\rangle_{F}}\\
&U_0=v_3^2\\
&U_1=2v_3^2+v_2^2\\
&U_2=6v_3^2+4v_2^2-2v_{2F}v_2\\
&U_3=1+15v_2^2-18v_{2F}v_2
\end{aligned}
\label{eq:c32p}
\end{equation}

The first term in the four-particle triangular cumulant coefficient, $c_3\{4\}|_p$, with a momentum $p$, reads
\begin{equation}
\begin{aligned}
&\langle e^{i3(\phi_1+\phi_2-\phi_3-\phi_4)}\rangle|_p=U_0+U_1Y+U_2\dfrac{Y^2}{2}+...+U_6\dfrac{Y^6}{720},
\quad ~Y=-\frac{p^2}{(N-4)\langle p^2\rangle_{F}}\\
&U_0=v_3^4\\
&U_1=4v_3^4+4v_2^2v_3^2\\
&U_2=28v_3^4+48v_2^2v_3^2+4v_2^4-8v_{2F}v_2v_3^2\\
&U_3=4v_3^2+256v_3^4+564v_2^2v_3^2+72v_2^4-264v_{2F}v_2v_3^2-48v_{2F}v_2^3\\
&U_4=160v_3^2+64v_2^2+6880v_2^2v_3^2+1056v_2^4-6224v_{2F}v_2v_3^2-1680v_{2F}v_2^3\\
&U_5=4180v_3^2+2400v_2^2+87044v_2^2v_3^2+14800v_2^4-800v_{2F}v_2-40800v_{2F}v_2^3\\
&U_6=400+89120v_3^2+59760v_2^2-42000v_{2F}v_2
\end{aligned}
\label{eq:U0U6appr}
\end{equation}

Using Eqs. (\ref{eq:cn2468}) and Eqs. (\ref{eq:c32p}-\ref{eq:U0U6appr}), we can get the final $c_3\{4\}|_p$. Note that in the approximate Eqs. (\ref{eq:c32p}) and (\ref{eq:U0U6appr}), we can not see any terms involving $\Psi_3$. It indicates that the effect of $\Psi_3$ is negligible, although the $\Psi_3$ terms do exists in the full results (see Appendix~\ref{appendix}).

\section{Application to $c_2\{4\}$, $c_3\{2\}$ and $c_3\{4\}$}
\label{compare}
In this section, we compare three observables ($c_2\{4\}$, $c_3\{2\}$ and $c_3\{4\}$)  with the experimental measurements,  and show the TMC and collective flow effects on them. For simplicity, we assume that all particles carry a common transverse momentum $p$, and we consider several reasonable values of $p$. The value of $\langle p^2\rangle_F$ is always taken to be $0.135$ (GeV/c)$^2$, and the flow parameters are chosen as follow: $v_2=0.05,\, v_3=0.0175,\, v_{2F}=0.025$, which are estimated based on the related experimental measurements. It should be pointed out that although we choose these parameters as some constants for simplicity, obviously they should depend on the multiplicity of $N$ in real experiments. In the following, we will present the results of $c_2\{4\}$, $c_3\{2\}$ and $c_3\{4\}$, which are obtained from full expressions as given in Appendix~\ref{appendix}.

\begin{figure}[h]
\begin{minipage}[t]{8cm}
\center
\includegraphics[scale = 0.3]{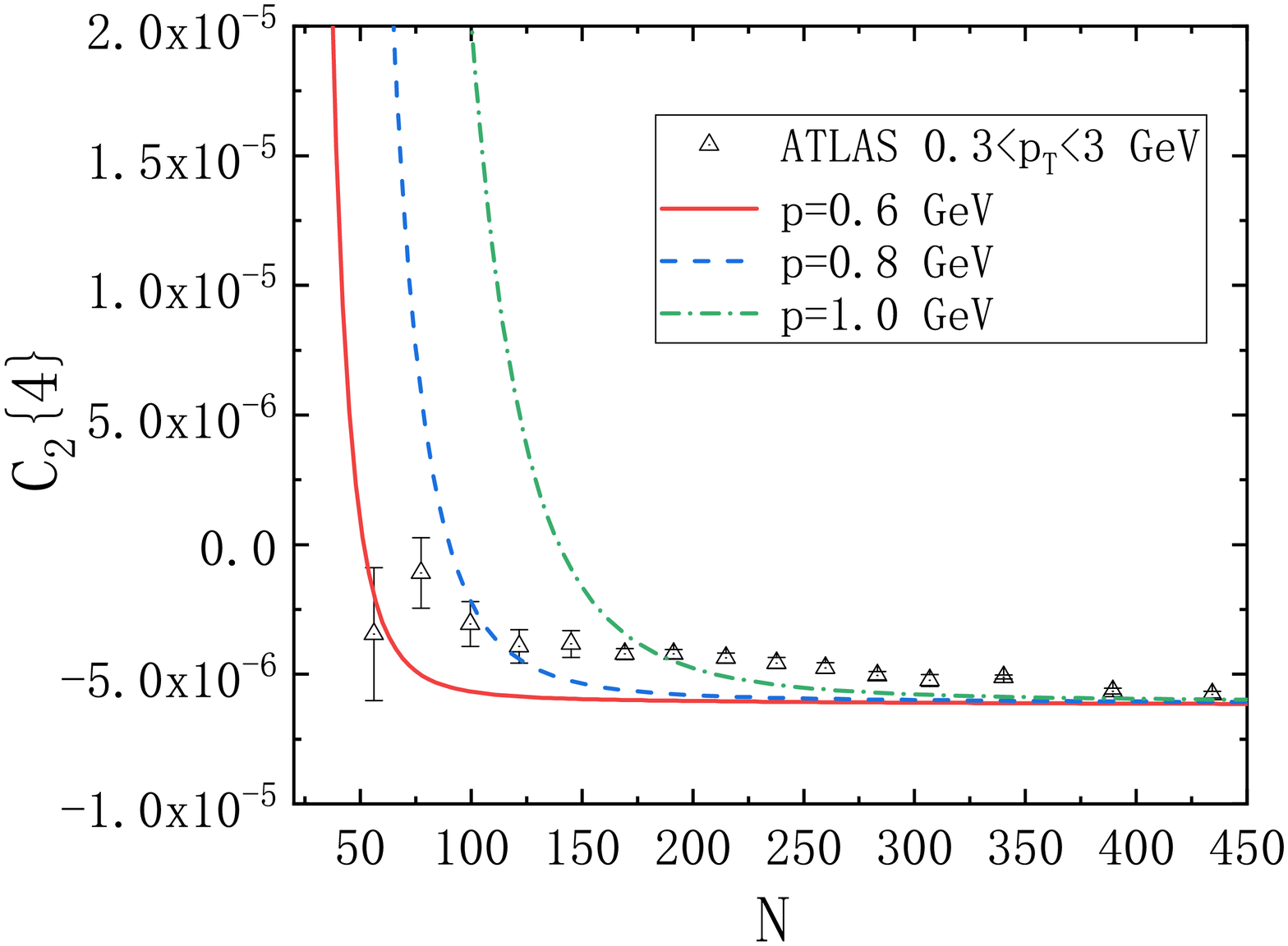}
\end{minipage}
\begin{minipage}[t]{8cm}
\center
\includegraphics[scale = 0.3]{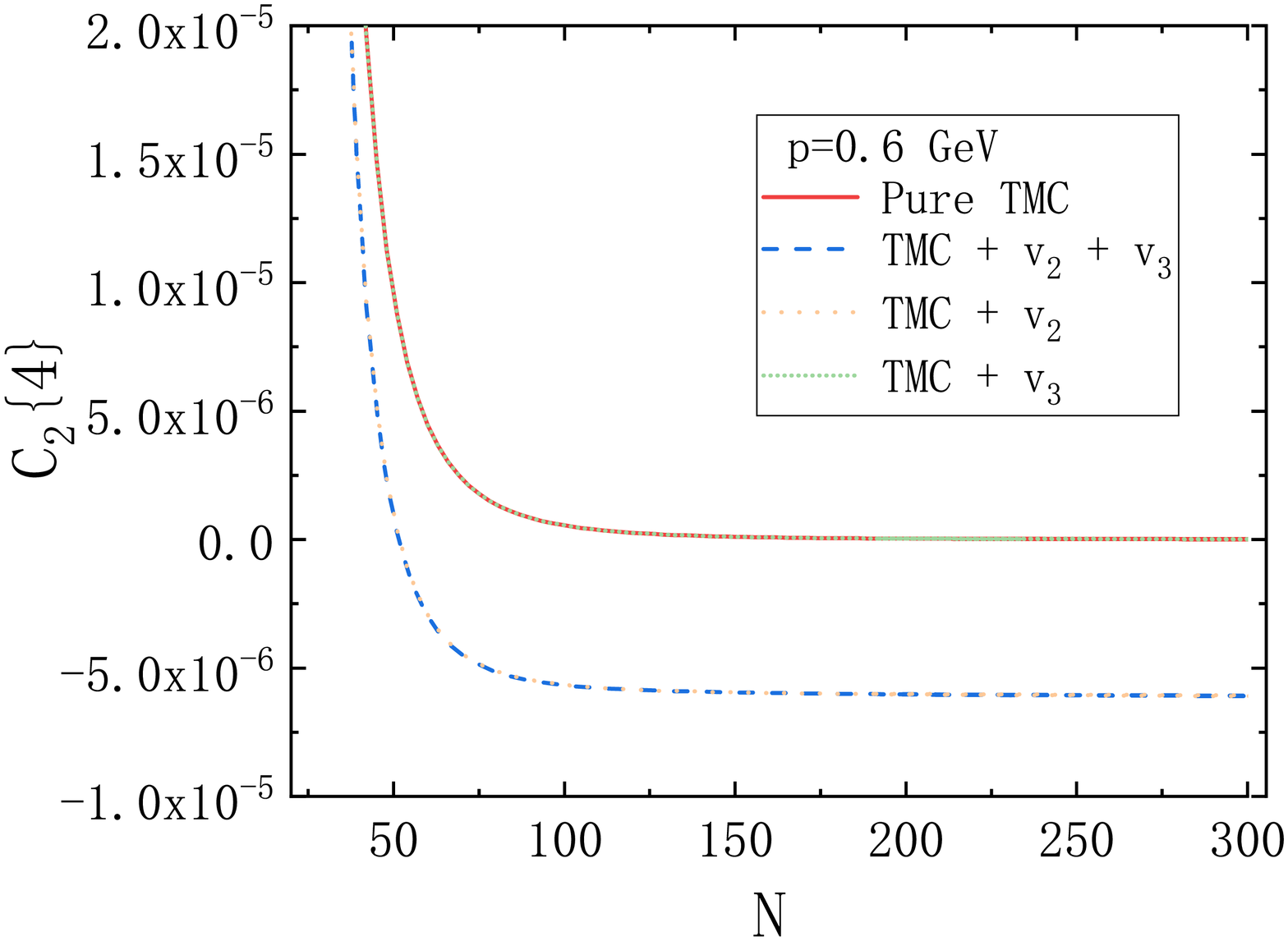}
\end{minipage}
\caption{(Left) $N$ dependence of $c_2\{4\}$ for the three selected momenta $p$, where 
the ATLAS data for p+Pb 5.02 TeV are shown for comparisons~\cite{Aaboud:2017blb}; (Right)  $N$ dependences of $c_2\{4\}$ from ``Pure TMC", ``TMC + $v_2$", ``TMC+$v_3$" and ``TMC+$v_2+v_3$", respectively, for the case of $p$=0.6 GeV. Note that the curve of ``Pure TMC"  almost overlaps with that of ``TMC+$v_3$", and the curve of ``TMC + $v_2$"  almost overlaps with that of ``TMC+$v_2+v_3$".}
\label{fig:C24}
\end{figure}

In the left plot of Fig.~\ref{fig:C24}, we show our results on the $N$ dependence of $c_2\{4\}$ including the TMC and collective flow  ($v_2$ and $v_3$) contributions for three selected momenta $p$, in comparisons with the ATLAS data for p+Pb 5.02 TeV~\cite{Aaboud:2017blb}. Note that because the $N$ should stand for the total number of particles affected by TMC which is not equivalent to the number of detected charged particles, the $N$ of experimental data points is multiplied by a factor of 1.5 to take the neutral particles into account. Our results show a decreasing  tendency with increasing $N$ which resembles the data qualitatively. We also find that $c_2\{4\}$ changes its sign at different $N$ for different $p$, which was observed in our previous study~\cite{Bzdak:2018web}.

To illustrate how the TMC and collective flow can influence $c_2\{4\}$, the right panel of Fig. \ref{fig:C24} shows four different cases for $p=0.6$ GeV which include four kinds of contribution combinations, such as from the TMC only (denoted as``Pure TMC"), TMC and elliptic flow (denoted as``TMC + $v_2$"), TMC and triangular flow (denoted as ``TMC+$v_3$") and TMC and collective flow (denoted as``TMC+$v_2+v_3$"). We can see that the TMC only leads to a decreasing tendency with increasing $N$, and if further taking into account the elliptic flow $v_2$ the curve will be lowered which results in a sign change of $c_2\{4\}$. However, the presence of triangular flow $v_3$ has a negligible effect on $c_2\{4\}$.
\begin{figure}[h]
\begin{minipage}[t]{8cm}
\center
\includegraphics[scale = 0.3]{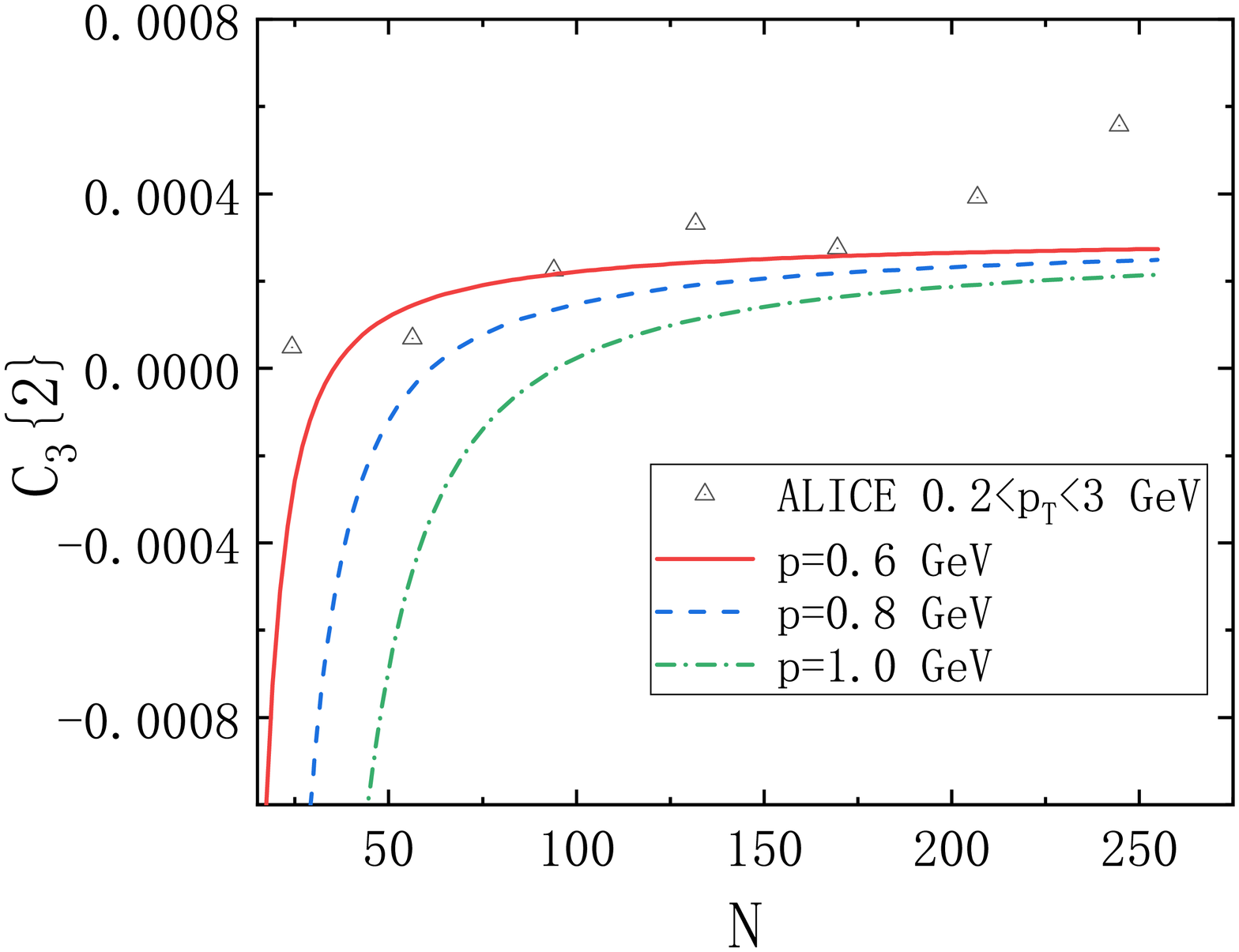}
\end{minipage}
\begin{minipage}[t]{8cm}
\center
\includegraphics[scale = 0.3]{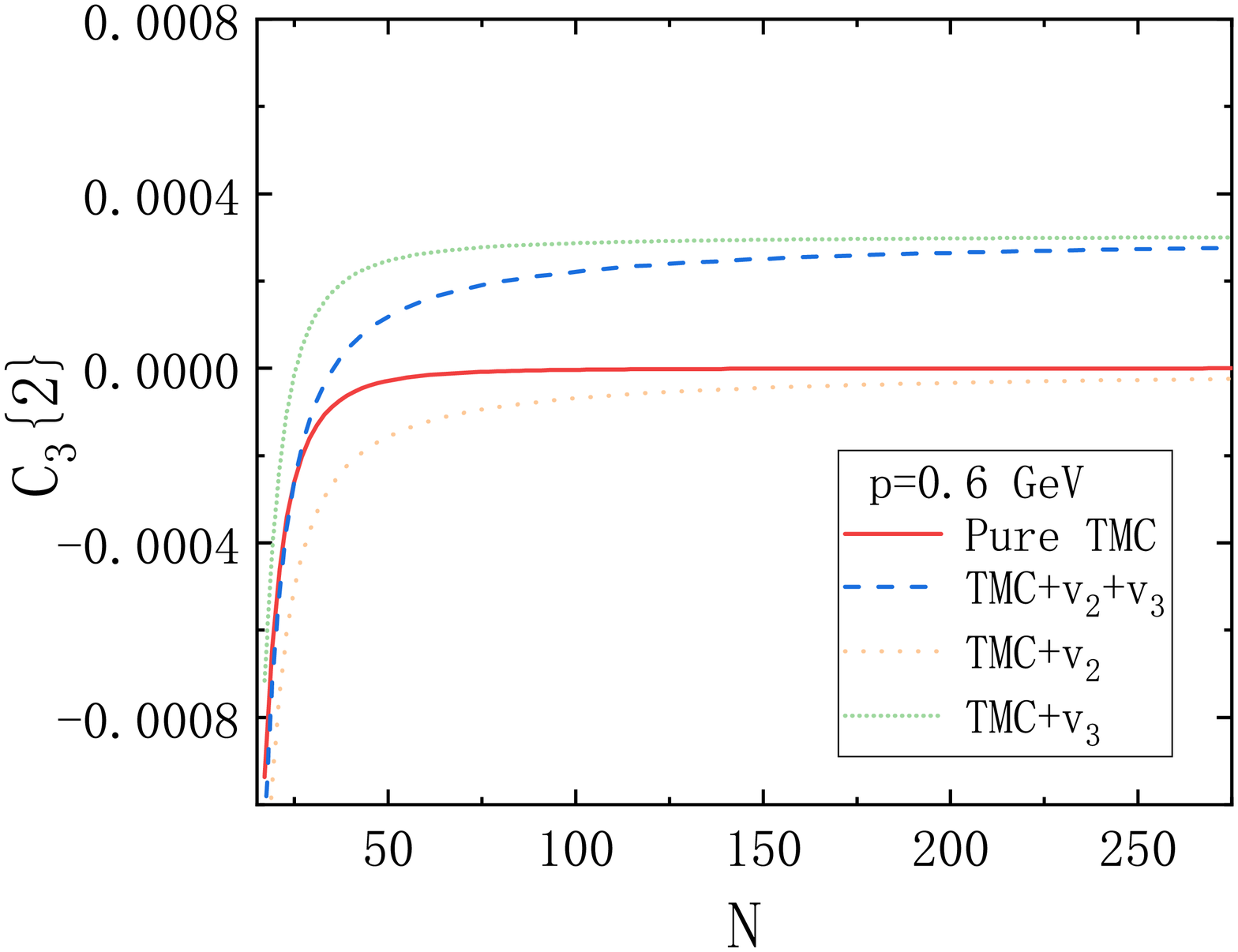}
\end{minipage}
\caption{Same as Fig.~\ref{fig:C24} but for $c_3\{2\}$, where the ALICE data for p+Pb collisions at 5.02 TeV are shown for comparison~\cite{Abelev:2014mda}.}
\label{fig:C32}
\end{figure}

The left plot of Fig.~\ref{fig:C32} shows our results on $N$ dependence of $c_3\{2\}$ due to the total effect of TMC and collective flow for three selected momenta $p$. They all show an increasing tendency with increasing $N$, which can describe the data qualitatively. In the right plot of Fig.~\ref{fig:C32}, we choose $p$=0.6 GeV to analyze different effects separately. As discussed in Table~\ref{table:TMC1}, the TMC effect results in a negative $c_3\{2\}$, which is our baseline to study any additional effect from collective flow. But $c_3\{2\}$ will be enhanced if only triangular flow $v_3$ exists, which can result in a sign change at a certain $N$. On the other hand, $c_3\{2\}$ becomes more negative if only elliptic flow $v_2$ is present. Therefore, we observe that $c_3\{2\}$ from ``TMC+$v_2$+$v_3$" is lower than that from ``TMC+$v_3$", because $v_2$ plays such a reducing role for $c_3\{2\}$.

\begin{figure}[h]
\begin{minipage}[t]{8cm}
\center
\includegraphics[scale = 0.3]{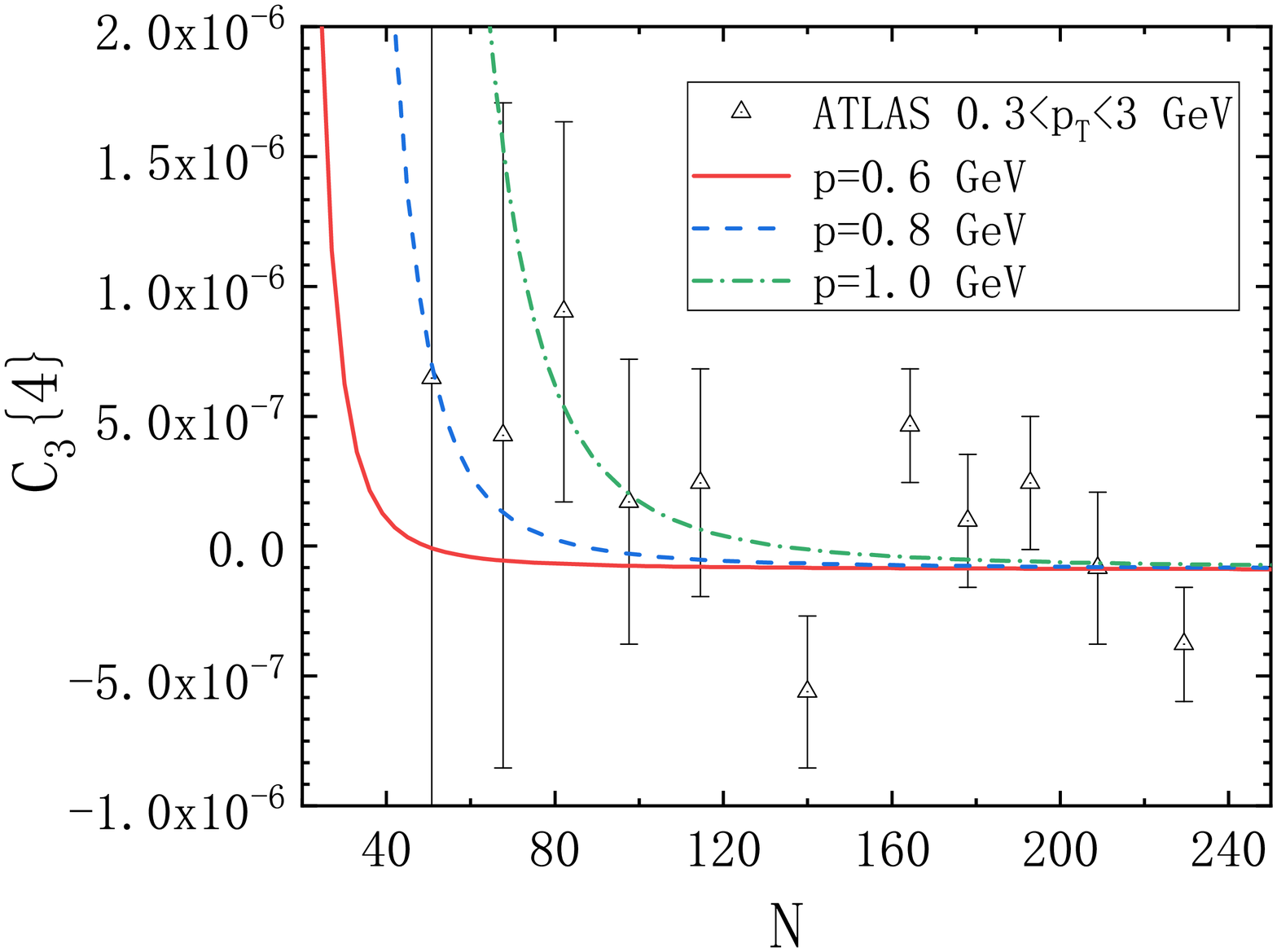}
\end{minipage}
\begin{minipage}[t]{8cm}
\center
\includegraphics[scale = 0.3]{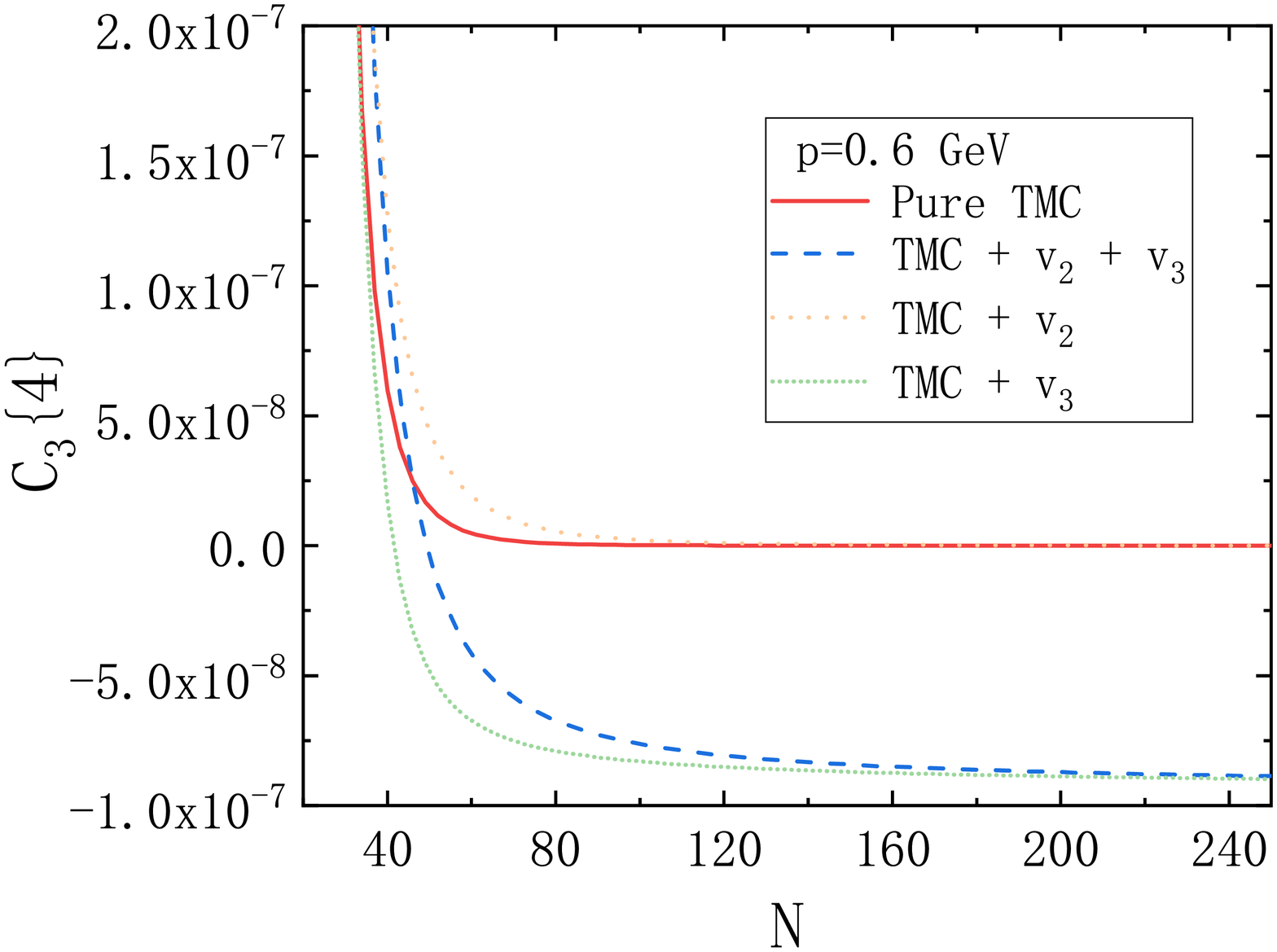}
\end{minipage}
\caption{Same as Fig.~\ref{fig:C24} but for $c_3\{4\}$, where the ATLAS data for p+Pb collisions at 5.02 TeV are shown for comparison~\cite{Aaboud:2017blb}.}
\label{fig:C34}
\end{figure}

In the left plot of Fig.~\ref{fig:C34}, we present the $N$ dependence of $c_3\{4\}$ due to the total effect of TMC and collective flow for three selected momenta $p$. We can observe a decreasing tendency of $c_3\{4\}$ with different magnitudes for different momenta $p$, which can describe the data qualitatively. Note that because the magnitude of $c_3\{4\}$ is so small that its sign change with increasing $N$ is hardly visible in the left plot. But this feature can be observed in the right plot of Fig.~\ref{fig:C34} with a smaller scale of the $y$-axis. By comparing the different kinds of cases, we see that although $v_2$ slightly raises $c_3\{4\}$ at small $N$, but $v_3$ significantly pushes $c_3\{4\}$ down which can result in a sign change of $c_3\{4\}$ with increasing $N$. Unfortunately, we can not see the sign change of $c_3\{4\}$ in the current experimental measurement due to large statistical uncertainties.  However, it is very helpful and important to measure the small sign change of $c_3\{4\}$ for both exploring the collectivity in small colliding systems and searching for the substructure of proton, because these triangular flow coefficients are expected to be more sensitive to the enhanced triangularity due to the existence of a three-hot spot substructure inside the proton~\cite{Mantysaari:2016ykx,Mantysaari:2020axf,Schenke:2021mxx,Zhao:2021bef}.

\section{conclusion}
\label{conclusions}
In this paper, we calculate the $n$th order of $2k$-particle azimuthal cumulant flow coefficients $c_n\{2k\}$ with the effects from transverse momentum conservation and collective flow (including both elliptic and triangular flow). We analytically demonstrate that the TMC only leads to a nonzero $c_n\{2k\}$ with the sign of $(-1)^{nk}$ and the magnitude inversely proportional to $(N-2k)^{nk}$. The results including both the TMC and collective flow are qualitatively comparable with the experimental measurements. We observe the sign changes of $c_3\{2\}$ and $c_3\{4\}$ with increasing multiplicity $N$ due to the interplay of TMC and collective flow, which could provide a good probe to study the onset of collectivity and the substructure of the proton in small colliding systems. We note that our analytic investigation should be viewed as the first approximation of the TMC effects. For example, we assumed that all momenta are fixed and identical. To address this and other issues one needs to investigate this problem using numerical methods. This would also allow obtaining a more precise result at a very small number of produced particles. The technique presented in this paper could be also used to study other correlations such as the four-particle symmetric cumulant $sc_{n,m}\{4\}$ and the three-particle asymmetric cumulant $ac_{n,m}\{3\}$. This could allow for more precise tests of the interplay of TMC and collective flow.

\section*{ACKNOWLEDGMENTS}
We thank Jean-Yves Ollitrault for useful correspondence. M.X. is grateful for the opportunity provided by the Fudan's Undergraduate Research Opportunities Program. G.M. is supported in part by the National Natural Science Foundation of China under Contracts No. 11961131011, No. 12147101, No. 11890710, No. 11890714, and No. 11835002, the Strategic Priority Research Program of Chinese Academy of Sciences under Grant No. XDB34030000, and the Guangdong Major Project of Basic and Applied Basic Research under Grant No. 2020B0301030008. AB was partially supported by the Ministry of Science and Higher Education, and by the National Science Centre, Grant No. 2018/30/Q/ST2/00101.

\appendix
\renewcommand{\appendixname}{Appendix}

\section{Calculation method}
\label{appendix}

When calculating $c_n\{2k\}$, every term coming from the expansion $(1+X+\frac{X^2}{2}+\frac{X^3}{6}+...)$ of exp(X) can be written as
\begin{equation}
C \exp{(i\sum a_j\phi_j)}
\end{equation}
of which the outcome of integration in the numerator of Eq.~(\ref{integ}) is determined by the power of each $e^{i\phi_j}$ and can be written as
\begin{equation}
C \prod v_{|b_j|}\exp(i \rm sgn(b_j)\Psi_{|b_j|}),
\end{equation}
where
\begin{equation}
b_j=\left\{
\begin{aligned}
n-a_j&, j\leq k \\
n+a_j&, j>k \\
\end{aligned}
\right.
\end{equation}
To illustrate the calculation, we first give a corresponding calculating table of integration outcome in TABLE~\ref{table:TMC2} and then an example.

\begin{table}[htbp]
\caption{The integration outcome for calculating $2k$-particle cumulant $c_n\{2k\}$. }
\label{table:TMC2}
\centering
\begin{tabular}{p{150pt}p{25pt}p{25pt}p{25pt}p{25pt}p{25pt}p{25pt}p{25pt}p{25pt}p{25pt}}
\hline
\hline
Initial term &1&$e^{i\phi_1}$&$e^{i\phi_2}$&$e^{-i\phi_1}$&$e^{-i\phi_2}$&$e^{2i\phi_1}$&$e^{2i\phi_2}$&$e^{3i\phi_1}$&$e^{3i\phi_2}$\\
\hline
Integration outcome ($n$=2)&$v_2^2$&$v_2v_3$&$v_1v_2$&$v_2v_3$&$v_1v_2$&$v_2v_4$&$v_2$&$v_2v_5$&$v_1v_2$\\
\\
Integration outcome ($n$=3)&$v_3^2$&$v_3v_4$&$v_2v_3$&$v_3v_4$&$v_2v_3$&$v_3v_5$&$v_1v_3$&$v_6v_3$&$v_3$\\
\hline
\hline
\end{tabular}
\end{table}

When calculating, e.g., $c_2\{2\}|_{p}$, the cumulant $c_2\{2\}|_p$ can be written as
\begin{equation}
c_2\{2\}|_p=U_0+U_1Y+U_2\dfrac{Y^2}{2},\quad Y=-\frac{p^2}{(N-2)\langle p^2\rangle_F(1-v_{2F}^2)}
\end{equation}
$U_0, U_1, U_2$ terms from the expansion $(1+X+\frac{X^2}{2})$ are obtained as follows.

Clearly, from the table above we can get the integral outcome of 1 is $U_0=v_2^2$.

The $U_1$ term from the expansion term $X$ can be first rewritten as
\begin{equation}
\begin{aligned}
X&=
(2e^0
-\dfrac{1}{2}v_{2F}e^{2i\phi_1}
-\dfrac{1}{2}v_{2F}e^{-2i\phi_1}
-\dfrac{1}{2}v_{2F}e^{2i\phi_2}
-\dfrac{1}{2}v_{2F}e^{-2i\phi_2}\\
&+e^{i(\phi_1-\phi_2)}
+e^{i(\phi_2-\phi_1)}
+v_{2F}e^{i(\phi_1+\phi_2)}
+v_{2F}e^{-i(\phi_1+\phi_2)})Y
\end{aligned}
\end{equation}
and thus the integration outcome is
\begin{equation}
\begin{aligned}
U_1&=
2v_2^2
-\dfrac{1}{2}v_{2F}v_2v_4e^{i(4\Psi_4-2\Psi_2)}
-\dfrac{1}{2}v_{2F}v_2e^{i(-2\Psi_2)}
-\dfrac{1}{2}v_{2F}v_2e^{i(2\Psi_2)}
-\dfrac{1}{2}v_{2F}v_2v_4e^{i(2\Psi_2-4\Psi_4)}\\
&+v_3^2
+v_1^2
+v_{2F}v_1v_3e^{i(3\Psi_3-\Psi_1)}
+v_{2F}v_1v_3e^{i(\Psi_1-3\Psi_3)},
\end{aligned}
\end{equation}
Neglecting all terms containing $v_1, v_4$ (they are assumed to be small) and putting $\Psi_2 = 0$, we obtain
\begin{equation}
U_1 = 2v_2^2+v_3^2-v_{2F}v_2
\end{equation}
Similarly, we have
\begin{equation}
U_2 = 1+6v_2^2+4v_3^2 - 8v_{2F}v_2+\dfrac{1}{2}v_{2F}^2(1+7v_2^2+4v_3^2)
\end{equation}

Following the same technique, $c_3\{2\}|_{p}$ is given by
\begin{equation}
c_3\{2\}|_p=U_0+U_1Y+U_2\dfrac{Y^2}{2}+U_3\dfrac{Y^3}{6}, \quad Y=-\frac{p^2}{(N-2)\langle p^2\rangle_F(1-v_{2F}^2)}
\end{equation}
\begin{equation}
\begin{aligned}
U_0&=v_3^2\\
U_1&=2v_{3}^{2}+v_{2}^{2}\\
U_2&=6v_{3}^{2}+4v_{2}^{2}-2v_{2F}v_{2}+3v_{2F}^{2}v_{3}^{2}+2v_{2F}^{2}v_{2}^{2}\\
U_3&=1+20v_3^2+15v_2^2-18v_{2F}v_2+\frac{3}{2}v_{2F}^2+30v_{2F}^2v_3^2+24v_{2F}^2v_2^2\\&-\frac{1}{4}v_{2F}^3v_3^2\cos(6\Psi_3)-\frac{9}{2}v_{2F}^3v_2
\end{aligned}
\end{equation}

For the term of $\langle e^{i2(\phi_1+\phi_2-\phi_3-\phi_4)}\rangle|_p$ in $c_2\{4\}|_{p}$ we have
\begin{equation}
\begin{aligned}
&\langle e^{i2(\phi_1+\phi_2-\phi_3-\phi_4)}\rangle|_p=U_0+U_1Y+U_2\dfrac{Y^2}{2}+U_3\dfrac{Y^3}{6}+U_4\dfrac{Y^4}{24},\quad Y=-\frac{p^2}{(N-4)\langle p^2\rangle_F(1-v_{2F}^2)}\\
&U_0=v_2^4\\
&U_1=4v_2^2v_3^2+4v_2^4-2v_{2F}v_2^2v_3^2\cos(6\Psi_3)-2v_{2F}v_2^3\\
&U_2=4v_3^4+4v_2^2+48v_2^2v_3^2+28v_2^4-24v_{2F}v_2v_3^2-36v_{2F}v_2^2v_3^2\cos(6\Psi_3)\\
&-40v_{2F}v_2^3+2v_{2F}^2v_3^4+12v_{2F}^2v_2v_3^2\cos(6\Psi_3)+5v_{2F}^2v_2^2+24v_{2F}^2v_2^2v_3^2+15v_{2F}^2v_2^4\\
&U_3=36v_3^2+72v_3^4+96v_2^2+564v_2^2v_3^2+256v_2^4-36v_{2F}v_2-672v_{2F}v_2v_3^2-564v_{2F}v_2^2v_3^2\cos(6\Psi_3)\\
&-702v_{2F}v_2^3+54v_{2F}^2v_3^2+108v_{2F}^2v_3^4+420v_{2F}^2v_2v_3^2\cos(6\Psi_3)+264v_{2F}^2v_2^2+879v_{2F}^2v_2^2v_3^2\\
&+432v_{2F}^2v_2^4-45v_{2F}^3v_3^2\cos(6\Psi_3)-9v_{2F}^3v_2-168v_{2F}^3v_2v_3^2-162v_{2F}^3v_2^2v_3^2\cos(6\Psi_3)\\
&-183v_{2F}^3v_2^3\\
&U_4=36+960v_3^2+1056v_3^4+1768v_2^2+6880v_2^2v_3^2+2720v_2^4-1440v_{2F}v_2-13968v_{2F}v_2v_3^2\\
&-8600v_{2F}v_2^2v_3^2\cos(6\Psi_3)-12016v_{2F}v_2^3+108v_{2F}^2+2880v_{2F}^2v_3^2+3168v_{2F}^2v_3^4\\
&+10332v_{2F}^2v_2v_3^2\cos(6\Psi_3)+8754v_{2F}^2v_2^2+22416v_{2F}^2v_2^2v_3^2+9732v_{2F}^2v_2^4\\
&-2100v_{2F}^3v_3^2\cos(6\Psi_3)-84v_{2F}^3v_3^4\cos(6\Psi_3)-1080v_{2F}^3v_2-10476v_{2F}^3v_2v_3^2\\
&-7822v_{2F}^3v_2^2v_3^2\cos(6\Psi_3)-9544v_{2F}^3v_2^3+\frac{27}{2}v_{2F}^4+360v_{2F}^4v_3^2+396v_{2F}^4v_3^4\\
&+2016v_{2F}^4v_2v_3^2\cos(6\Psi_3)+1238v_{2F}^4v_2^2+2876v_{2F}^4v_2^2v_3^2+\frac{5163}{4}v_{2F}^4v_2^4
\end{aligned}
\end{equation}
For the term of $\langle e^{3i(\phi_1+\phi_2-\phi_3-\phi_4)}\rangle|_p$ in $c_3\{4\}|_{p}$ we obtain
\begin{equation}
\begin{aligned}
&\langle e^{i3(\phi_1+\phi_2-\phi_3-\phi_4)}\rangle|_P=U_0+U_1Y+U_2\dfrac{Y^2}{2}+...+U_6\dfrac{Y^6}{720},
\quad Y=-\frac{p^2}{(N-4)\langle p^2\rangle_F(1-v_{2F}^2)}\\
&U_0=v_3^4\\
&U_1=4v_3^4+4v_2^2v_3^2-2v_{2F}v_2^2v_3^2\cos(6\Psi_3)\\
&U_2=28v_3^4+48v_2^2v_3^2+4v_2^4-8v_{2F}v_2v_3^2-36v_{2F}v_2^2v_3^2\cos(6\Psi_3)+14v_{2F}^2v_3^4\\
&+4v_{2F}^2v_2v_3^2\cos(6\Psi_3)+24v_{2F}^2v_2^2v_3^2+2v_{2F}^2v_2^4\\
&U_3=4v_3^2+256v_3^4+564v_2^2v_3^2+72v_2^4-264v_{2F}v_2v_3^2-564v_{2F}v_2^2v_3^2\cos(6\Psi_3)-48v_{2F}v_2^3\\
&+6v_{2F}^2v_3^2+384v_{2F}^2v_3^4+165v_{2F}^2v_2v_3^2\cos(6\Psi_3)+852v_{2F}^2v_2^2v_3^2+108v_{2F}^2v_2^4\\
&-5v_{2F}^3v_3^2\cos(6\Psi_3)-\frac{1}{2}v_{2F}^3v_3^4\cos(6\Psi_3)-66v_{2F}^3v_2v_3^2-144v_{2F}^3v_2^2v_3^2\cos(6\Psi_3)-12v_{2F}^3v_2^3\\
&U_4=160v_3^2+2716v_3^4+64v_2^2+6880v_2^2v_3^2+1056v_2^4-6224v_{2F}v_2v_3^2-8600v_{2F}v_2^2v_3^2\cos(6\Psi_3)\\
&-1680v_{2F}v_2^3+480v_{2F}^2v_3^2+8148v_{2F}^2v_3^4+4656v_{2F}^2v_2v_3^2\cos(6\Psi_3)+312v_{2F}^2v_2^2+21036v_{2F}^2v_2^2v_3^2\\
&+3240v_{2F}^2v_2^4-350v_{2F}^3v_3^2\cos(6\Psi_3)-44v_{2F}^3v_3^4\cos(6\Psi_3)-4668v_{2F}^3v_2v_3^2\\
&-6702v_{2F}^3v_2^2v_3^2\cos(6\Psi_3)-1260v_{2F}^3v_2^3+60v_{2F}^4v_3^2+\frac{2037}{2}v_{2F}^4v_3^4+790v_{2F}^4v_2v_3^2\cos(6\Psi_3)\\
&+44v_{2F}^4v_2^2+2646v_{2F}^4v_2^2v_3^2+408v_{2F}^4v_2^4\\
&U_5=4180v_3^2+31504v_3^4+2400v_2^2+87044v_2^2v_3^2+14800v_2^4-800v_{2F}v_2-127320v_{2F}v_2v_3^2\\
&-130560v_{2F}v_2^2v_3^2\cos(6\Psi_3)-40800v_{2F}v_2^3+20900v_{2F}^2v_3^2+157520v_{2F}^2v_3^4+110460v_{2F}^2v_2v_3^2\cos(6\Psi_3)\\
&+18300v_{2F}^2v_2^2+451300v_{2F}^2v_2^2v_3^2+78200v_{2F}^2v_2^4-14630v_{2F}^3v_3^2\cos(6\Psi_3)-2345v_{2F}^3v_3^4\cos(6\Psi_3)\\
&-1200v_{2F}^3v_2-190980v_{2F}^3v_2v_3^2-208360v_{2F}^3v_2^2v_3^2\cos(6\Psi_3)-61760v_{2F}^3v_2^3+\frac{15675}{2}v_{2F}^4v_3^2\\
&+59070v_{2F}^4v_3^4+\frac{113925}{2}v_{2F}^4v_2v_3^2\cos(6\Psi_3)+7650v_{2F}^4v_2^2+\frac{342495}{2}v_{2F}^4v_2^2v_3^2+29850v_{2F}^4v_2^4\\
&-\frac{7315}{4}v_{2F}^5v_3^2\cos(6\Psi_3)-\frac{2345}{8}v_{2F}^5v_3^4\cos(6\Psi_3)-100v_{2F}^5v_2-15915v_{2F}^5v_2v_3^2-\frac{71603}{4}v_{2F}^5v_2^2v_3^2\cos(6\Psi_3)-5170v_{2F}^5v_2^3\\
&U_6=400+89120v_3^2+387140v_3^4+59760v_2^2+1134720v_2^2v_3^2+206064v_2^4-42000v_{2F}v_2-2417184v_{2F}v_2v_3^2\\
&-1985256v_{2F}v_2^2v_3^2\cos(6\Psi_3)-856800v_{2F}v_2^3+3000v_{2F}^2+668400v_{2F}^2v_3^2+2903550v_{2F}^2v_3^4\\
&+2374920v_{2F}^2v_2v_3^2\cos(6\Psi_3)+662400v_{2F}^2v_2^2+9019260v_{2F}^2v_2^2v_3^2+1698120v_{2F}^2v_2^4\\
&-469140v_{2F}^3v_3^2\cos(6\Psi_3)-95140v_{2F}^3v_3^4\cos(6\Psi_3)-105000v_{2F}^3v_2-6042960v_{2F}^3v_2v_3^2\\
&-5433750v_{2F}^3v_2^2v_3^2\cos(6\Psi_3)-2184840v_{2F}^3v_2^3+2250v_{2F}^4+501300v_{2F}^4v_3^2+\frac{4355325}{2}v_{2F}^4v_3^4\\
&+2489940v_{2F}^4v_2v_3^2\cos(6\Psi_3)+550350v_{2F}^4v_2^2+6891660v_{2F}^4v_2^2v_3^2+1312695v_{2F}^4v_2^4\\
&-\frac{351855}{2}v_{2F}^5v_3^2\cos(6\Psi_3)-\frac{71355}{2}v_{2F}^5v_3^4\cos(6\Psi_3)-26250v_{2F}^5v_2-1510740v_{2F}^5v_2v_3^2\\
&-\frac{5681529}{4}v_{2F}^5v_2^2v_3^2\cos(6\Psi_3)-551565v_{2F}^5v_2^3+125v_{2F}^6+27850v_{2F}^6v_3^2+\frac{483925}{4}v_{2F}^6v_3^4\\
&+\frac{231}{8}v_{2F}^6v_3^4\cos(12\Psi_3)+\frac{319869}{2}v_{2F}^6v_2v_3^2\cos(6\Psi_3)+\frac{64125}{2}v_{2F}^6v_2^2+\frac{1545615}{4}v_{2F}^6v_2^2v_3^2+\frac{148059}{2}v_{2F}^6v_2^4\\
\end{aligned}
\end{equation}

\end{document}